# Revealing Peri-Urban Dislocation through Percolation Analysis

Dr Melissa Barrientos-Trinanes[1,2], Professor Stephen Marshall[1], Professor Elsa Arcaute[2]

[1] *The Bartlett School of Planning, University College London, London, UK*
[2] *Centre for Advanced Spatial Analysis (CASA), University College London, London, UK*

**Corresponding author:**
Dr Melissa Barrientos-Trinanes
melissa.barrientos.15@ucl.ac.uk

**ORCIDs:**
Dr Melissa Barrientos-Trinanes: 0000-0002-5885-6392
Professor Elsa Arcaute: 0000-0001-6579-3265
Professor Stephen Marshall: 0000-0003-3654-3835

**Abstract**

This paper introduces peri-urban dislocation as a structural condition that complements existing sprawl metrics by capturing hierarchical misalignments between inner-city and peripheral areas. Whereas conventional measures emphasise density, land-use mix, or fragmentation, peri-urban dislocation reflects deeper divergences in the core–periphery relational functional organisation of urban systems. We operationalise this concept using percolation analysis of street networks, revealing hierarchical patterns via clustering maps and dendrograms, providing a relational structure between urban elements. Two case studies, Valdivia, Chile, and Boston, USA, demonstrate contrasting manifestations: a structural reversal in Valdivia, where a homogeneous residential periphery dominates the hierarchical clustering process, and peri-urban voids in Boston, where isolated parcellations persist despite metropolitan consolidation. These findings position peri-urban dislocation as a structural dimension linked to sprawl yet distinct from metrics based on density or peripherality; one that may occur independently or represent a previously unidentified structural signature of sprawl. Methodologically, we apply established percolation techniques to expose this previously unarticulated structural phenomenon, enabling the detection of hierarchical misalignments within urban systems. Conceptually, we introduce *peri-urban dislocation* as a new dimension of urban structure, helping articulate debates on sprawl and peri-urbanisation through a complexity-informed lens and enabling core-periphery diagnostics across diverse urban contexts.

## 1. Introduction

Urban systems exhibit complexity through hierarchical coherence and hidden regularities and repetitions across scales. These characteristics, often described as fractal-like structures, are central to understanding cities as complex systems (Batty, 2008). As such, cities are in a constant state of adaptation (Durack, 2001; Jensen & Arcaute, 2010), with expansions often added to the periphery in a fractal-like manner.

Newly planned urban projects frequently exhibit less initial complexity than inner-city areas, adopting structures closer to Alexander's mathematical trees than to overlapping semi-lattices (Alexander, 2013). Yet, over time, this difference typically diminishes as new elements, such as small businesses, schools, or transit infrastructure, are introduced. However, some designed areas retain the homogeneity of their original spatial configuration and land uses, failing to adapt or participate in the hierarchical coherence observed in the broader urban system. This warrants attention, as persistent homogeneity has been linked to urban segregation, diminished vitality, dysfunction, inert milieux, socio-economic challenges, and the delayed formation of new local centres (Baldwin Hess et al., 2018; Hillier, 1996; Jacobs, 1961; Karimi & Vaughan, 2017). This may reflect a structural gap in the integration of peripheral growth into the urban system, or hierarchical misalignments between inner-city and peripheral areas. Such misalignments often manifest as disconnected zones that fail to adapt or develop centralities, reinforcing segregation and limiting functional integration within the urban system.

In hierarchically coherent urban systems, peripheral expansions integrate progressively with the urban core, forming nested structures where elements at different scales connect in an organised manner: smaller neighbourhoods linking to larger districts, which in turn connect to metropolitan centres (Batty, 2008; Pumain, 2006). This hierarchical organisation reflects how urban elements cluster and relate across spatial scales, from local streets to regional networks (Alexander, 2013; Arcaute et al., 2016). When peripheral areas develop in structural alignment with this hierarchy, they gradually integrate into the broader system through progressively stronger connections. However, hierarchical misalignments occur when peripheral expansions cluster independently, forming internally connected zones that remain structurally disconnected from the core, lacking the progressive integration that characterises coherent urban development.

Expanding peripheries shaped by mid-20th-century mass-produced developments often retain the uniformity of their original design. In addition, they might also lack functional adaptability, struggling to develop centralities over time (Lehmann, 2019), thereby maintaining a challenging spatial segregation. This creates a significant contrast with traditional inner cities shaped through layered planned and unplanned processes. Although these areas may appear related to sprawl, their disparity extends beyond location and density to morphological and structural dimensions. This is a form of spatial differentiation that reflects the designed nature of specific peripheral environments and that has not been systematically captured in the existing literature. We introduce the concept of *peri-urban dislocation* to describe this phenomenon: structural misalignments between the spatial logics of inner-city and peri-urban areas that suggest either a new structural dimension of sprawl or a distinct, independent phenomenon. This research identifies and quantifies these structural dislocations, using methods from complexity science.

We employ percolation clustering on a graph representation of the street network, using intersection distances as in studies by Arcaute et al. (2016), Cao et al. (2020), Fluschnik et al. (2016), and Marin et al. (2023). This approach reveals hierarchical patterns in the clustering process, exposing the underlying structure of the urban system. By identifying clustering and hierarchical structuring across the city, including discontinuities in the process, we apply this approach to detect, in a novel way, sudden changes in the transition between peri-urban areas and their inner-city counterparts that could reflect dislocation. By analysing how urban elements relate to one another in a spatial hierarchy, using percolation analysis to investigate these structures, we can detect shifts in the expected hierarchical organisation. These can manifest either as alterations in core-to-periphery clustering or as reversals in which peripheral areas dominate the clustering process, producing a periphery-to-core structure. Such structural changes have not been explicitly captured by conventional sprawl metrics. Percolation analysis provides a systematic means to quantify them, offering a structural indicator that complements existing density-based and morphological measures of sprawl.

## 2. Peri-urban and inner-city structural relations

### 2.1. A gap in sprawl measurement

Urban sprawl is commonly understood as low-density, spatially dispersed expansion from the urban core into peripheral or rural areas. Although traditionally defined in terms of density, the concept remains ambiguous, hindering consensus on measurement (Chettry, 2023; Bosch et al., 2019). Yet, existing indices have made significant advances in capturing multiple dimensions of sprawl. Multidimensional approaches incorporate density, land-use mix, accessibility, and socio-economic factors (Galster et al., 2001; Ewing et al., 2002; Simon, 2023; Chettry, 2023), while structural indices address fragmentation, clustering, centrality, nuclearity, and proximity (Galster et al., 2001; Ewing et al., 2002; Frenkel & Ashkenazi, 2008; Hamidi et al., 2015; Yue et al., 2016; Imbrenda et al., 2022). Clustering analysis has been used to delineate peri-urban zones (Goerlich Gisbert et al., 2017), and methods such as Shannon's entropy and discontinuity indices assess spatial and temporal dynamics (Feng et al., 2016) and the dispersion and concentration of built-up areas (Chettry, 2023). Other work integrates centrality and density indicators (Zhou et al., 2019), while fractal and landscape ecology measures quantify fragmentation and land-use diversity (Bosch et al., 2019). Remote sensing and GIS remain central tools for peri-urban analysis alongside proximity and flow models, Cellular Automata, and machine learning (Sahana et al., 2023).

However, these metrics largely measure sprawl characteristics within peripheral zones, examining density gradients, land-use patterns, and fragmentation as properties of the periphery, independent of the core. While these approaches successfully identify sprawl, given by spatial and morphological attributes in peripheral areas, they do not reveal how peripheral expansions integrate, or fail to integrate, within the hierarchical structure of the broader urban system. The question of structural coherence between core and periphery remains largely unaddressed. Most research focuses primarily on outer areas or on comparative analyses among peripheral cases, overlooking core-to-periphery relational dynamics. Authors such as Bosch et al. (2019) and Artmann et al. (2018) have emphasised that structural changes and the relationship between inner and outer cities remain underexplored. Studies have failed to comprehensively analyse the hierarchical transitions between urban, peri-urban, and rural regions (Cattivelli, 2021; Hoffmann et al., 2017; Mustak et al., 2018; Xia et al., 2020). Comparative studies often remain scale-specific and rarely examine how peripheral expansions relate structurally to pre-existing cores, generating a growing call for multi-scalar approaches (Artmann et al., 2019). As noted by Sahana et al. (2023), a deeper understanding of peri-urban complexity requires examining relationships, dynamics, systems, and internal/external distinctions. This gap is significant because it obscures whether peripheral areas develop as structurally integrated extensions of the urban system or as functionally disconnected zones that cluster independently and lack meaningful hierarchical ties to the core. To address this gap, we propose a quantitative and systematic framework to reveal hierarchical patterns and structural coherence, or their absence, between the core and the periphery.

### 2.2. Peri-urban dislocation

To articulate these observed contrasts in how urban elements relate, which at times resemble urban sprawl but are not fully so, we introduce the concept of *peri-urban dislocation*. Peri-urban dislocation describes a structural condition observed at a single point in time; a spatial configuration, rather than a chronological account of urban development. It refers to an observed structural variation pattern in which peri-urban areas appear as a shift from the traditional urban core and form well-defined clusters within the hierarchical structure at different distances from those at which a clearly defined core emerges. In this configuration, peri-urban elements cluster together more closely than they integrate with the broader system. However, despite their internal cohesion, they lack a functional core of their own. Dislocation implies segregation from the main urban system without the formation of autonomous hubs, often resulting in dormitory-like areas. To classify an area as a dislocation, we suggest that three conditions must be met:

(1) Independent clustering within the structural model.
(2) No immediate connection to the core of the city.
(3) Absence of a strong urban centre or centrality with diverse land uses and activity hubs.

A peri-urban dislocation captures structural misalignments that conventional sprawl metrics overlook, offering a new lens for understanding peri-urban fragmentation beyond density and location. However, while structural differences between urban areas are common, not all of them constitute a dislocation. If a contrasting area develops a strong core or centrality, it remains a structural difference rather than a dislocation. Dislocation occurs only when the area lacks such a core, reinforcing its segregation. This distinction clarifies why dislocation is more than a morphological contrast; it represents a systemic misalignment within the hierarchical organisation leading to functional effects (see later Figure 5).

## 3. Methodology

### 3.1. Clustering methods

Clustering methods are widely used to identify homogeneous aggregates within heterogeneous urban environments. Techniques such as K-means, DBSCAN, and percolation clustering have been applied to detect grouping patterns in spatial data. Previous studies have used clustering to delineate peri-urban zones, classify urban form, and explore structural characteristics (Sahana et al., 2023; Goerlich Gisbert et al., 2017). While these approaches provide valuable insights, they do not capture relationships among urban elements such as those between inner and outer areas, which can be analysed using clustering methods that yield a hierarchical structure.

Urban morphology studies have also used clustering for classification, but limitations persist regarding scalability, transferability, robustness, extensiveness, and interpretative flexibility (Fleischmann et al., 2022). Similarly, clustering techniques have been applied to urban morphology metrics using elements such as building footprints. While comprehensive, these approaches depend on the availability and consistency of input data, which can be problematic, particularly for building footprints. In contrast, street network data is generally more reliable, consistent, and widely available across contexts. The street network also serves as a strong proxy for identifying highly planned environments and the spatial organisation embedded in the urban form. For these reasons, our analysis focuses on the street network and the spatial distribution of POIs as the primary observational element.

### 3.2. Percolation and hierarchical analysis

Percolation clustering goes one step further than conventional clustering. It reveals the hierarchical organisation of the clustering process, thereby exposing the system's underlying structure rather than merely the distances or densities among elements. It also detects any structural breaks. Within a network-based approach, it enables exploration of connectivity across scales, from local clusters to entire metropolitan systems, without requiring a pre-specified number of clusters. Percolation is a data-driven process in which the number of clusters need not be specified in advance, unlike other clustering algorithms. This is particularly well-suited to identifying the underlying standard structure of the spatial distribution, as well as any unexpected breaks that deviate from it and could indicate a dislocation.

Percolation has been applied to uncover latent spatial hierarchies and clustering dynamics in road networks at metropolitan and national scales (Arcaute et al. 2016; Cao et al. 2020; Fluschnik et al. 2016; Marin et al. 2023). It has been used at the city scale to reveal the hierarchical structure of the street network and other urban elements (Piovani et al., 2017; Huynh et al., 2018; Hu et al., 2022; Kwon et al., 2023). Specifically for urban sprawl, it has been used to examine the connectivity and density of buildings (Behnisch et al., 2019). Yet, it has not been implemented to explore the structural contrast between inner-city and peri-urban areas. Its capacity to examine network connectivity while recovering the hierarchical organisation of urban systems makes it particularly well-suited to identifying peripheral developments that diverge from traditional organisations. This paper extends percolation analysis to detect structural breaks in the core-to-periphery transition and explores peripheral growth through the lens of structural contrast rather than solely through density, function, or location.

We apply percolation clustering to a primal graph of the street network (intersections as nodes; segments as edges; see Porta et al., 2006), following the methodology established in previous studies. Distance thresholds (d_t) increase in 10 m increments from 10 m until full network connectivity (typically ≥ 500 m). At each d_t, intersections form clusters that progressively merge within the analytical process, revealing a hierarchical organisation and structural breaks. We report three outputs: (i) clustering maps, showing spatial aggregation across d_t; (ii) LCC, or Largest Connected Component charts, tracking the growth of the largest cluster in the model and highlighting abrupt expansions that indicate discontinuities in cluster growth; and (iii) dendrograms, or hierarchical trees, visualising hierarchical aggregation and branch depth. Taken together, these three tools provide complementary insights to track cluster growth patterns at each distance threshold, allowing us to observe how the network structure changes as connectivity increases.

We focus on identifying discontinuities in the clustering process and evaluating whether these correspond to peri-urban dislocations. We considered three structural indicators: (i) independent clustering at low and high thresholds, indicating peripheral areas that remain spatially separated from the core across the percolation process; (ii) abrupt changes in LCC growth that do not correspond to physical discontinuities or administrative schemes, suggesting unexpected structural breaks; and (iii) changes in the depth of branching patterns in dendrograms, revealing shifts in hierarchical integration.

### 3.3. Data sources

We used open-source geospatial street network data, extracted from OpenStreetMap using OSMnx (Boeing, 2017) in Python. Percolation clustering, dendrogram generation, and visualisation were performed in RStudio using scripts adapted from Arcaute et al. (2016). This approach enabled us to examine the structural organisation across scales, detect discontinuities, and identify peri-urban dislocations in line with the criteria outlined earlier.

### 3.4. Case studies

We apply this methodology to two highly suburbanised traditional cities: Valdivia, Chile, and Boston, USA. These cases offer contrasting yet complementary insights into mid-20th-century urban expansion in the Americas. Valdivia was selected because of rapid, real estate-led peripheral growth following neoliberal reforms from the 1970s, resulting in master-planned, mass-produced housing developments creating a stark morphological contrast with the traditional city. Boston was selected for its low-density peri-urban growth linked to post-war suburban parcelling, embedded within a complex suburban landscape including both recent developments and mature suburbs dating to the 1600s. Boston's extended suburbanisation timeline allows observation of structural misalignments as residual effects of earlier sprawl.

### 3.5. An analytical framework for identifying dislocation

Percolation methods, previously applied to identify hierarchical clustering in UK road networks and reveal socio-political and economic regional structures (Arcaute et al., 2016), offer a quantifiable way to reveal underlying spatial structures within cities. We use the three outputs of percolation in combination to reveal the hierarchical structure and identify dislocations in urban systems. Clustering maps are used to expose which intersections connect at each distance threshold, visualising the evolution of spatial clusters. Largest Connected Component (LCC) charts track the dominant cluster's size across thresholds, with abrupt increases (or "jumps") indicating discontinuities. Dendrograms display the hierarchical tree structure, showing how and when clusters merge. Together, these outputs indicate whether clustering follows expected hierarchical patterns or exhibits structural breaks that suggest dislocation. We highlight the ten largest clusters at each threshold: the largest in red, the second and third in blue and green, and the remaining seven in other colours. This colour-coding tracks how clusters merge and which areas dominate at different scales.

In terms of expected patterns, in traditional urban systems, clustering seems to follow a core-first logic, with the urban core forming the dominant cluster at lower distance thresholds, gradually absorbing peripheral areas as thresholds increase. The pattern can be recognised in clustering maps produced by Piovani et al. (2017) and Arcaute et al. (2016) for cities such as London, Birmingham, Manchester, Liverpool, and Bristol. Urban systems may show multiple early clusters, but secondary clusters typically integrate with the primary core as connectivity increases.

Regarding structural discontinuities, even conventional hierarchies exhibit discontinuities when geographical or built barriers (rivers, highways, parks) are overcome. We term abrupt, disproportionate increases in the largest cluster as "jumps." These reflect expected discontinuities as clustering overcomes physical barriers and typically align with administrative boundaries or the known urban structure.

The percolation analysis operationalises the three conditions from Section 2.2: (1) Independent clustering manifests when peripheral areas form dominant clusters at low thresholds rather than the core, reversing the expected sequence. (2) Lack of connection to the core appears as persistent independent clustering across thresholds, with peripheral areas forming independent dendrogram branches rather than integrating with the main hierarchical tree. (3) Abrupt breaks unrelated to physical barriers or secondary centres indicate unexpected discontinuities that distinguish dislocation from polycentric development.

While such developments often exhibit sprawl characteristics (low density, automobile dependence), the structural disconnect revealed through percolation represents a dimension that conventional sprawl metrics do not capture, namely, whether peripheral growth integrates hierarchically with the urban core or develops as a structurally independent system without forming coherent centres.

## 4. Results

### 4.1. Valdivia

#### 4.1.1. Expected clustering pattern

Valdivia's urban development has unfolded in four phases (Espinoza Guzmán et al., 2016). The city originated in the 16th century and remained a small town until the mid-19th century (Phase I). During the industrial era (late 19th–early 20th century, Phase II), new neighbourhoods emerged independently on either side of large rivers, creating today's historic districts: *Guacamayo, Las Animas, Isla Teja, Collico,* and *Torobayo*. These two phases constitute the traditional city, shaped by a mix of organic and planned growth, with Phase I forming the foundational core and Phase II comprising river-separated areas.

Since Phase III (the 1960s), Valdivia has doubled in size, with approximately 65% of its current area developed through top-down planning. The post-1960 earthquake reconstruction triggered rapid peri-urban expansion to provide social housing, followed by Phase IV: accelerated suburbanisation from the 1970s onward, primarily through market-led master-planned, mass-produced housing. This pattern of suburban expansion is consistent with the characteristics of urban sprawl, particularly in its low-amenity peripheral growth, single-family housing focus, and large-scale residential development (Galster et al., 2001; Ewing et al., 2002).

Valdivia's expected pattern is core-first, with historic neighbourhoods branching independently in the dendrogram until merging with the core at higher distance thresholds. Given the city's rivers and wetlands, we anticipate discontinuities where these historic neighbourhoods cluster separately prior to integrating at higher distance thresholds.

#### 4.1.2. Structural breaks

The percolation analysis shown in Figure 1 reveals a significant structural shift at a distance threshold of 130 m, marking a turning point in Valdivia's clustering dynamics. Two distinct regimes are observed. Above 130 m, the largest cluster, corresponding to the foundational centre and urban core, expands gradually and then exhibits pronounced jumps, highlighted in pink in the LCC chart (Figure 1a). These jumps reflect the absorption of historic neighbourhoods separated by rivers, representing expected discontinuities caused by natural barriers. In the dendrogram (Figure 1b), these neighbourhoods appear as long, independent branches that remain separate for many levels until merging with the central cluster. This pattern reflects the city's historical growth, in which *Guacamayo, Las Animas, Isla Teja, Collico*, and *Torobayo* developed independently and joined the core at higher thresholds. We labelled these as 'historic branches.'

Below 130 m d_t, however, marked with a black square and an 'x', the clustering pattern changes dramatically. Between 70 m and 130 m, the largest cluster depicted in red grows in a cascade, highlighted in cyan (Figure 1a). This cluster is only formed of peripheral areas, excluding the urban core in blue (Figure 1a). This peripheral clustering contrasts with the expected core-first one and signals a structural reversal. The dendrogram confirms in Figure 1b this shift. Instead of a deep hierarchical tree linked to the core and branching outward, the structure shows shallow branches, indicating that peri-urban areas cluster at lower distances, dominating the system. The foundational core appears at 100–120 m and merges into the central cluster at 130 m (highlighted in a blue transition). Below this threshold, the periphery consolidates independently from the core, disrupting the conventional hierarchy.

#### 4.1.3. A structural reversal

Valdivia's peri-urban dislocation manifests as a structural reversal and a dominance of the periphery. The clustering pattern below 130 m reveals a striking turnaround of the expected urban hierarchy (Figure 1b). Instead of the foundational core forming the largest cluster first, peri-urban areas dominate the structure at lower thresholds. By 110 m, the largest cluster no longer includes the traditional core; it consists almost entirely of peripheral developments. This shift is driven by post-1960s expansions, characterised by master-planned housing schemes with highly homogeneous street networks. These areas cluster uniformly between 70 m and 120 m, forming a single dominant cluster without strong local centres. The dendrogram reflects this pattern: shallow branches proliferate from one threshold to another at lower distances, indicating an aggregation of peripheral nodes, while the core emerges at higher distances and remains structurally secondary (Figure 1b). This contrasts sharply with the conventional core-first logic, shifting to an unexpected periphery-to-core sequence that signals a structural dislocation within the urban system.

Beyond the extreme homogeneity of street layouts in the periphery, stemming from large-scale, highly planned developments built from scratch, the landscape itself was transformed. Wetlands that once constrained development were replaced by landfills, eliminating the natural discontinuities that had historically structured the city's growth. As a result, peripheral clusters are internally cohesive yet disconnected from the traditional core, lacking the hierarchical depth and diversity that characterise the historic city. This structural misalignment is not a temporary phase but a systemic departure from Valdivia's historical growth pattern.

If we had mapped the city using only the traditional core and historic neighbourhoods, excluding post-1960s development, the dendrogram would have shown a different development pattern. It would have followed the historic logic: the foundational core forming at lower distance thresholds, at the bottom of the dendrogram, and remaining the primary cluster as distance increased. The peripheral consolidation observed between 70 m and 100 m would not exist. Instead, the system would have maintained its core-first structure, gradually incorporating the independent historic branches separated by rivers.

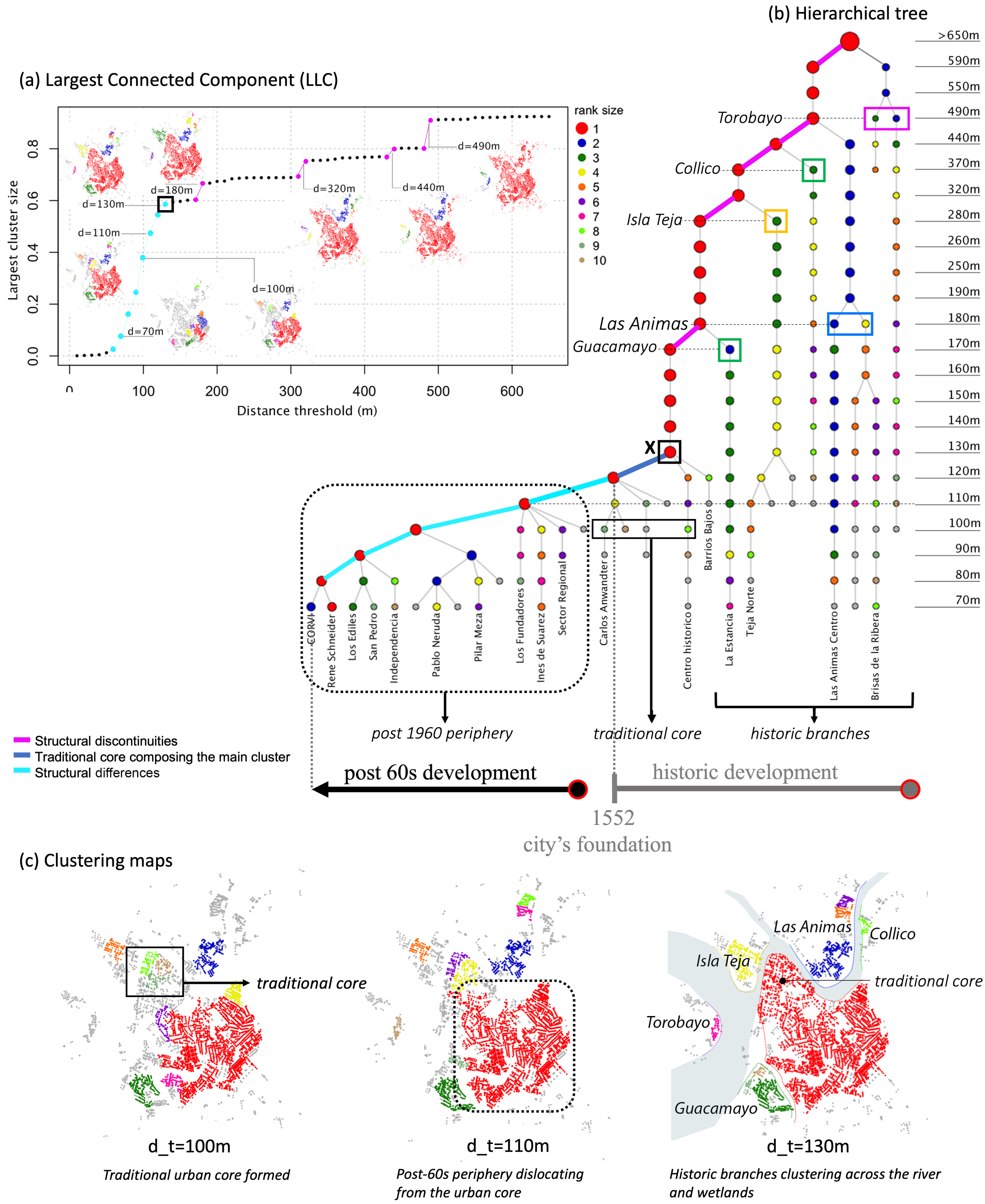

**Figure 1. Hierarchical structure and structural reversal in Valdivia.** (a) LCC chart showing cluster growth. Pink highlights mark the absorption of historic neighbourhoods. Cyan highlights the 70–130 m cascade, where the periphery consolidates at lower thresholds than the core. (b) Dendrogram illustrating the formation of the foundational core at 100–120 m (blue) and the historic neighbourhoods appearing as long branches. Below 130 m, marked as 'X' and with a black square, the post-1960 periphery clusters at lower thresholds than the core, dominating the system and reversing the expected core-first pattern that would otherwise have emerged at 100 m with the traditional core. (c) Clustering maps showing a periphery-first dominance and a core-integrated clustering.

#### 4.1.4. Implications

We have uncovered a structural difference between the periphery and the traditional city, revealing a dislocation that goes beyond the typical structural discontinuities the city has exhibited historically. This is evident in the hierarchical clustering process, which highlights areas that form independent structural units and alter the established growth pattern.

Valdivia's small urban core makes the city particularly vulnerable to the dominance of peripheral clusters. Analysis of land-use data confirms that the periphery lacks strong centralities or local centres (Figure 2), with post-1960 developments showing predominantly residential land uses, without the concentration of services, commercial activity, or institutional functions that would indicate strong secondary centres. This absence of strong centralities, combined with independent clustering, constitutes a dislocation as these areas diverge structurally from the central city. As the residential periphery expands, it risks overshadowing the traditional city in both size and structural influence, creating large dormitory suburbs that remain functionally dependent on the core without developing autonomous centres. This imbalance undermines the formation of new centralities and perpetuates spatial segregation, limiting opportunities for mixed-use development and local economic vitality. The structural reversal revealed by percolation analysis shows that peri-urban dislocation is not a transient anomaly, but a systemic condition embedded in the city's growth logic since the post-1960s. These findings highlight the need for planning strategies that go beyond density-based metrics to address structural coherence, ensuring that future expansions integrate hierarchical diversity and connectivity rather than reinforcing homogeneity and isolation.

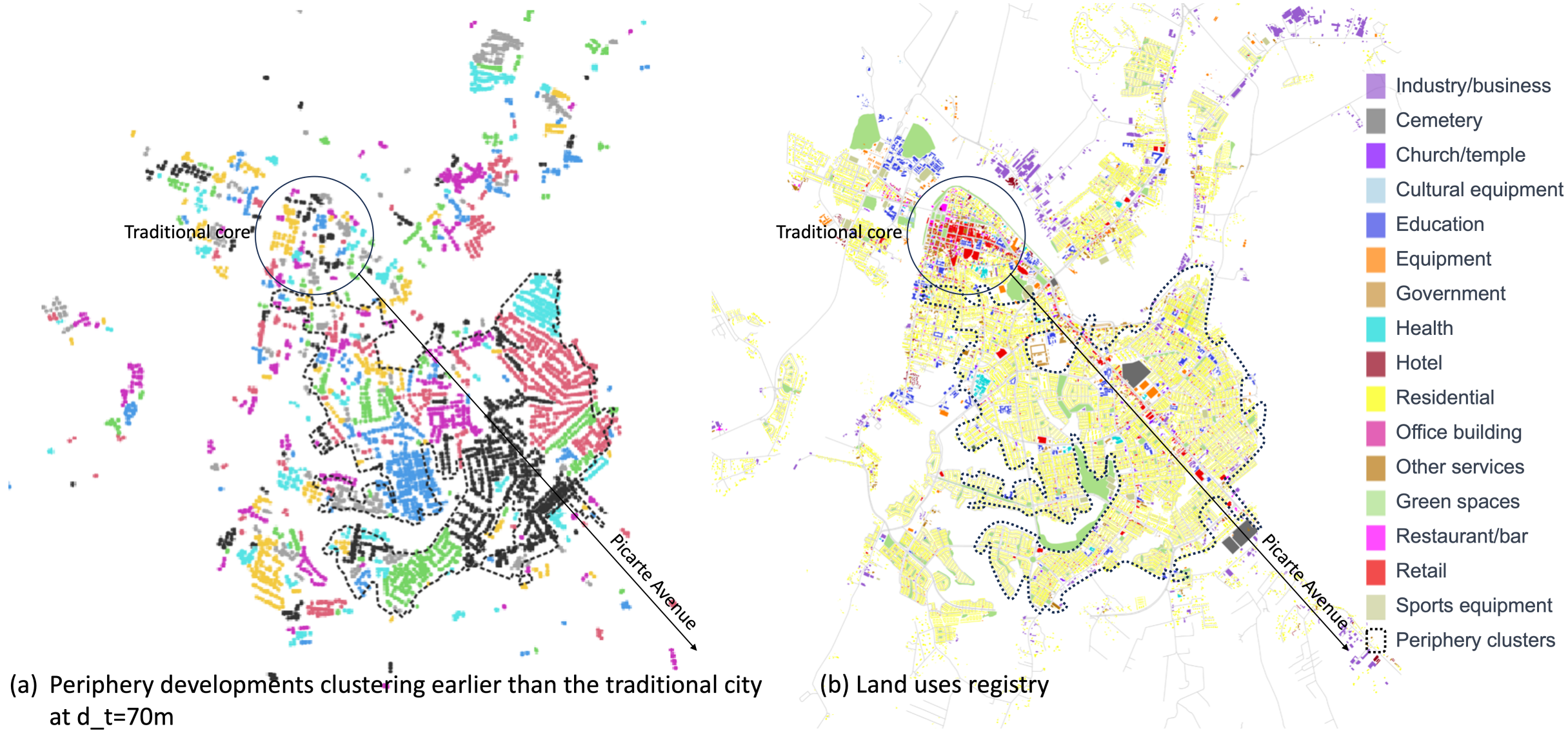

(a) Periphery developments clustering earlier than the traditional city at d_t=70m
(b) Land uses registry

**Figure 2.** Evidence of absent centralities in Valdivia's periphery. (a) The colours represent the different clusters at d_t=70m, a threshold at which the uniform street structure of peripheral developments produces tight, consistent neighbourhood clusters, while the more heterogeneous intersections of the traditional core have not yet consolidated into a single cluster. (b) A plot of land-use data confirms that the periphery lacks a functional core: no independent strong concentration of services, commercial activity, or institutional functions. The only visible non-residential corridor is Picarte Avenue, a linear extension of the city core rather than an autonomous secondary centre. Post-1960 developments are predominantly residential.

## 4.2. Boston

### 4.2.1. Expected clustering pattern

Boston's metropolitan area, defined by the Metropolitan Area Planning Council (MAPC), comprises 101 municipalities with Boston and Cambridge at its core. Within this region, the Inner Core Communities (ICC) include Boston and 21 surrounding municipalities. Historically, Boston developed as a dense city, later expanding through suburbanisation and regional integration. Its post-war suburban expansion is widely interpreted as a form of urban sprawl, characterised by low-density residential growth, automobile dependence, and extensive single-family zoning (O'Connell, 2013). As with Valdivia, the expected clustering pattern for such a system is core-first, with the central cities of Boston and Cambridge clustering independently and then merging, as soon as the Charles River is overcome, together with some immediate neighbours, such as Charlestown, to form the largest cluster at lower thresholds. This would be followed by the gradual absorption of peripheral areas as distance increases until the MAPC is complete. As in Valdivia, we anticipate finding structural discontinuities primarily along the Charles River and highways. Administrative boundaries in urban systems often emerge from historical settlement patterns, infrastructure development, and natural barriers. If this holds for Boston, clustering thresholds might correspond, to some degree, to administrative divisions (neighbourhoods → cities → ICC → MAPC).

#### 4.2.2. Structural breaks

Percolation analysis confirms the expected core-to-periphery logic but reveals notable discontinuities aligned principally with the river and administrative boundaries. The first recognisable clusters appear at 60 m, corresponding to neighbourhoods concentrated in Boston and Cambridge. At 110 m, Boston, Cambridge, and Charlestown emerge as distinct clusters (Figure 3c). At 140 m, Boston and Cambridge cross the Charles River and merge, consolidating the historic core. The ICC completes at 190 m, while the MAPC boundary is reached at 360 m. These transitions correspond to regular jumps in the LCC chart (Figure 3a), highlighted in pink, and reflect the expected discontinuities.

However, two anomalies disrupt this otherwise conventional core-to-periphery pattern. First, an unexpected northward expansion beyond the MAPC boundary occurs between 300 and 320 m: at d_t=300m, the system is still clustering within the MAPC boundary, but d_t=320m shows the main cluster in red expanding outside the metropolitan boundary to absorb areas to the north, marked by sharp jumps on the LCC chart (in cyan, Figure 3, c.2.). This is anomalous because it incorporates areas that fall outside the study boundary before full metropolitan integration is achieved. Second, within the MAPC, persistent peri-urban voids, defined here as independent peripheral clusters that lack centralities and fail to achieve structural integration with the core, remain disconnected from the surroundings even after metropolitan consolidation. At 360 m, 42 isolated zones, falling within the segregated areas inside dotted line rectangles shown at d_t=360 in Figure 3, c.2, remain unabsorbed; even at 410 m, they remain detached from the central cluster. These voids correspond to *Developing Suburbs*, low-density areas with fragmented connectivity, as highlighted by dotted rectangles on the maps (Figure 3c, 3d).

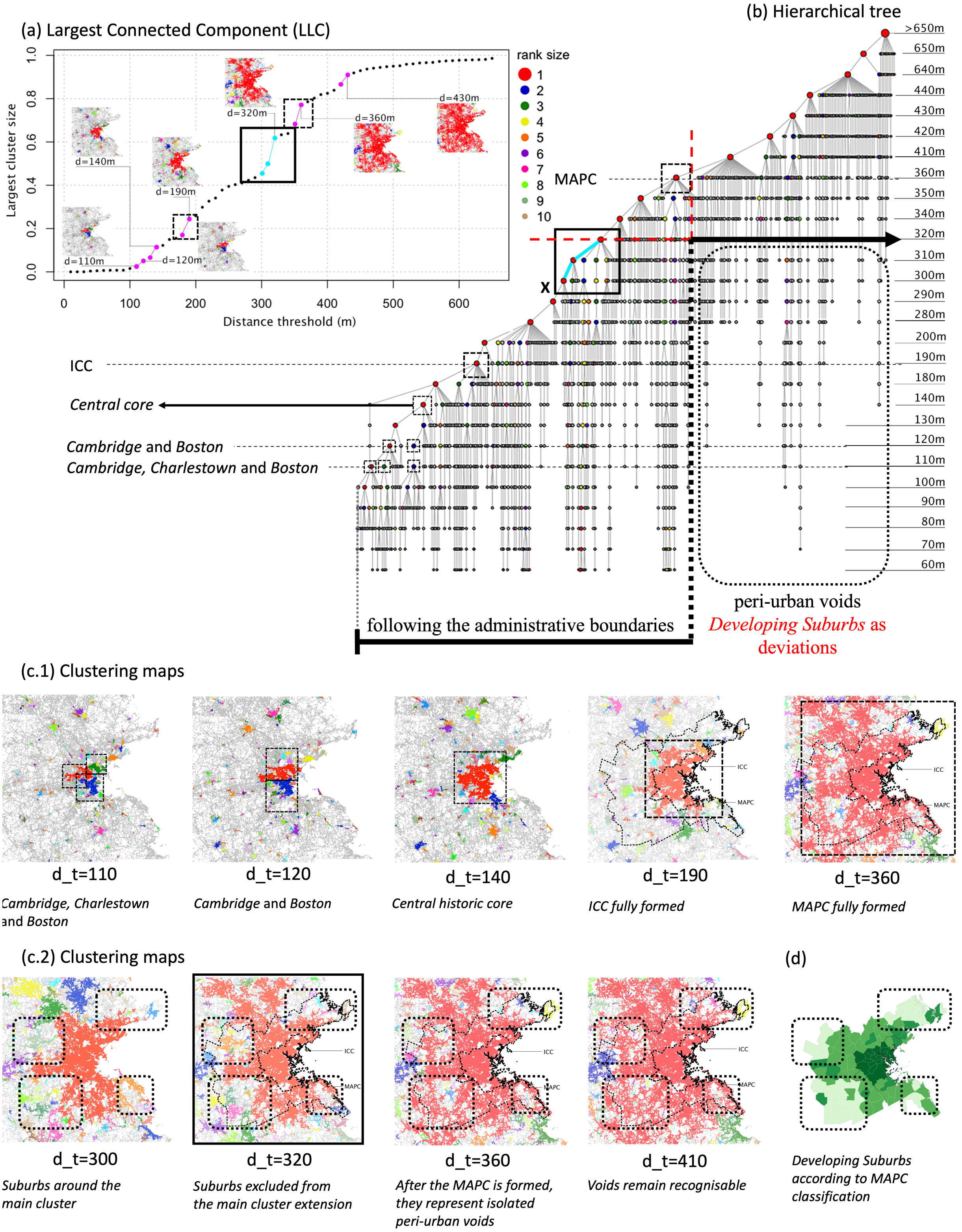

**Figure 3. Hierarchical structure from percolation clustering in Greater Boston.** (a) LCC chart showing cluster growth across thresholds, with regular jumps aligned to administrative boundaries (pink) and unexpected jumps at 300–320 m (cyan, marked as an 'x' rectangle in Fig. 3.b). (b) Dendrogram illustrating hierarchical aggregation: Boston, Cambridge, and Charlestown form at 110 m; ICC completes at 190 m; MAPC consolidates at 360 m. Peri-urban voids remain isolated beyond 410 m. (c) Clustering maps at selected thresholds, showing Boston–Cambridge consolidation, ICC formation, northward expansion beyond MAPC, and isolated Developing Suburbs (dotted rectangles).

### 4.2.3. Conventional hierarchy with peri-urban voids

Peri-urban dislocation in Boston manifests as persistent voids within an otherwise conventional core-to-periphery hierarchy. Unlike Valdivia, Boston does not exhibit a reversal of hierarchy; the core remains dominant throughout the clustering process. The system exhibits structural gaps and residual internal dislocations. The clustering continues to extend beyond the metropolitan boundary, while areas inside it remain isolated as peri-urban voids, signalling structural misalignments within an otherwise integrated system. In the dendrogram (Figure 3b), these voids materialise at higher levels (70 m, 100 m, and 280 m), in contrast to clusters aligned with administrative boundaries (mostly at 60 m). Their isolation indicates that certain peripheral areas fail to integrate structurally, even as the metropolitan system consolidates. These anomalies may represent residual effects of earlier suburbanisation patterns, producing gaps in the urban fabric that conventional sprawl metrics overlook.

### 4.2.4. Implications

Boston's case demonstrates that peri-urban dislocation can manifest as structural voids within an otherwise hierarchical system, unlike the hierarchical reversal observed in Valdivia. While the metropolitan system retains a core-to-periphery logic, concentrated gaps undermine structural coherence, perpetuating fragmented suburban landscapes. The mapping of land-use data confirms the absence of strong centralities in these peripheral voids (Figure 4b), with isolated areas exhibiting predominantly residential uses and lacking the concentration of services, commercial activity, or institutional functions that would indicate strong secondary centres at a metropolitan level. These findings underscore the need for planning approaches that address connectivity and integration at the metropolitan scale, ensuring that peripheral zones achieve structural integration regardless of their distance from the core.

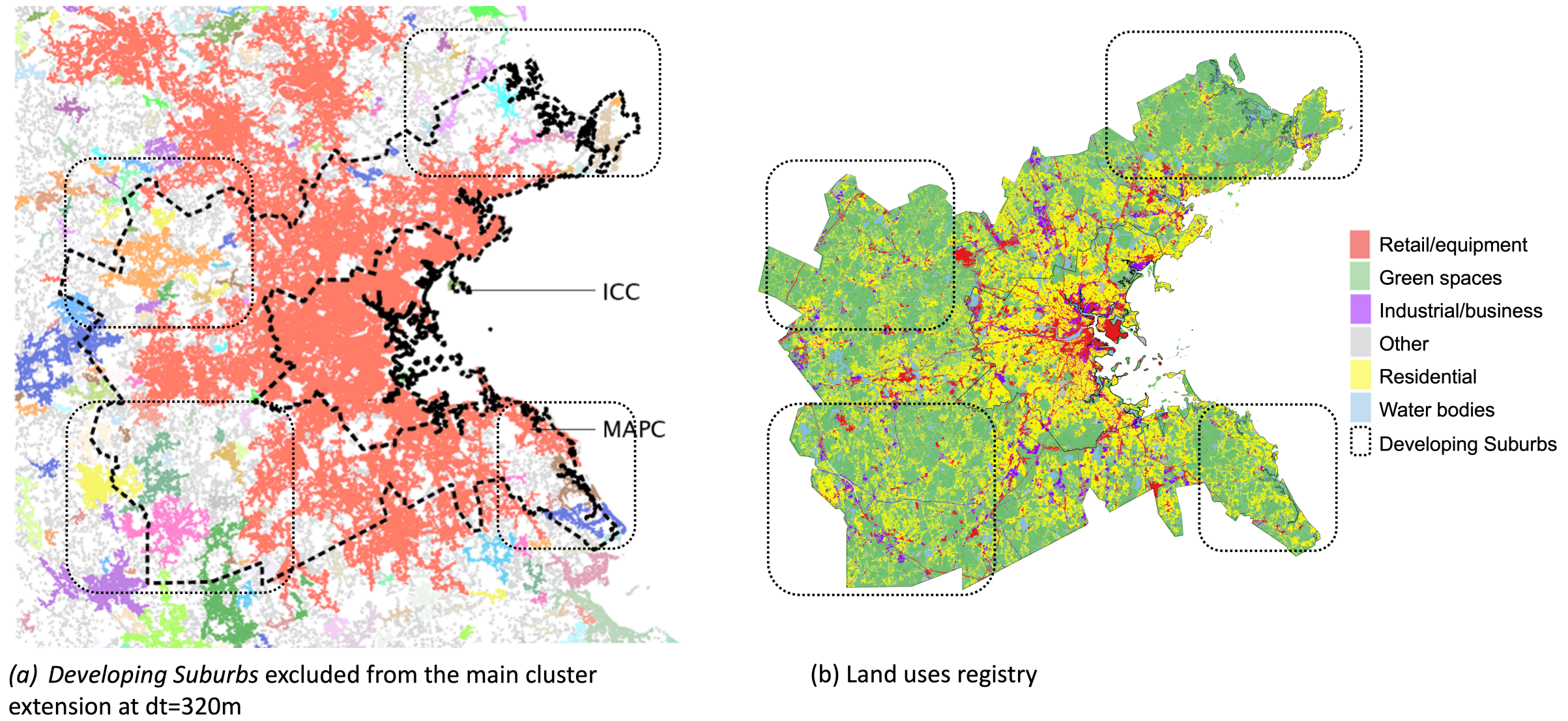

**Figure 4.** Evidence of the absence of strong centralities in Boston's peri-urban voids. (a) A map showcasing the isolated peripheral suburbs that remain disconnected even after metropolitan consolidation. (b) Land-use distribution confirms the absence of strong functional cores in these areas: predominantly residential, with concentrations of services, commercial activity, or institutional functions insufficient to constitute robust, independent centres at the metropolitan scale.

## 5. Discussion

### 5.1. Comparative insights

Together, these cases illustrate two distinct expressions of peri-urban dislocation: one driven by rapid, planned expansions following the neoliberal reforms of the 1970s, which reversed traditional urban hierarchies (Valdivia, Figure 1), and another shaped by the long-term effects of post-war suburban parcelling (Boston, Figure 3). Despite their different trajectories, both reveal structural disparities between peri-urban development and the inner-city system.

5.2. Conceptualising peri-urban dislocation

We define peri-urban dislocation as a structural variation in which peripheral areas (i) cluster independently within the hierarchical process, (ii) do not connect directly to the city's core, and (iii) lack strong centralities, meaning an absence of diverse land-use agglomeration or activity hubs that would constitute secondary centres. Mapping of land-use data in both Valdivia and Boston confirms the absence of strong centralities in peripheral areas (Figures 2 and 4). While dislocations can occur as part of sprawl, they are not reducible to low density, car orientation, or peripherality; they can also arise in higher-density layouts when relational organisation diverges from the traditional matrix.

We identify two interpretive scenarios for the conceptualisation of peri-urban dislocation and its relationship to sprawl:

1) A remnant of sprawl (persistent structural misalignments from the urban fabric of the inner city to a new structure accompanying peripheral expansion, that remain after subsequent urban consolidation, once areas are no longer peripheral and are embedded in later growth);
2) A distinct structural phenomenon, emerging independently of sprawl, as a misfit in the system's hierarchical coherence.

Valdivia aligns mainly with (2) because of the reversal in a periphery that is not necessarily low-density relative to the system. On the other hand, Boston mostly exemplifies (1), with residual voids within an otherwise integrated metro structure, but still in a low-density fashion. Our cases suggest that dislocation can be well understood through these scenarios, as they capture the phenomenon's persistence beyond sprawl (Boston) and its emergence independent of sprawl characteristics (Valdivia).

5.3. Methodological reflection

Percolation clustering, assessed through three complementary outputs, LCC charts, dendrograms, and clustering maps, reveals structural misalignments beyond what morphology alone can capture. In Valdivia, dendrogram depth and branching factor revealed a qualitative change: the traditional city's deep, single-child branches were consistent with historical growth along rivers and wetlands, whereas shallow, wide peripheral branches, forming at shorter distances unlike elsewhere, characterised the periphery. In Boston, the LCC was pivotal in diagnosing regular jumps that correspond broadly to administrative boundaries (Boston–Cambridge merger at 140 m; ICC completion at 190 m; MAPC consolidation at 360 m) and, critically, to unexpected jumps beyond the metropolitan boundary at 300–320 m. Peri-urban voids persist as disconnected fragments (42 isolated clusters at 360 m; still detached at 410 m). This persistent isolation distinguishes voids from areas experiencing a slower integration.

Using these three tools together enables quantitative, reproducible identification of dislocations by exploring the system's hierarchy. This approach operationalises principles of complexity science for urban diagnostics, bridging network theory and spatial planning.

5.4. Manifestations and typology

Dislocation is best understood against a baseline of traditional urban organisation (core-first, multi-scalar coherence). We observe:

I) Type I: A traditional city formed through layered planned/unplanned growth. It clusters from the core, including non-dislocated peripheral suburbs that are structurally coherent with the core and/or possess local centres that mitigate misalignment.
II) Type II: Peri-urban dislocation, with two subtypes:
   - Sub-type IIa: a peripheral reversal**:** homogeneous planned units coalescing at lower distances in a large, central cluster, overpowering the traditional core; a *difference in kind* that re-orders the hierarchical process from periphery-to-core.
   - Sub-type IIb: peri-urban voids: independent clusters lacking centralities and remaining detached at metropolitan scales; a *difference in degree* within a core-first hierarchy.

Our empirical analysis demonstrates the two dislocation subtypes (Figure 5). Type IIa is observed in the Valdivia peripheral reversal. Type IIb is seen in Boston's peri-urban voids.

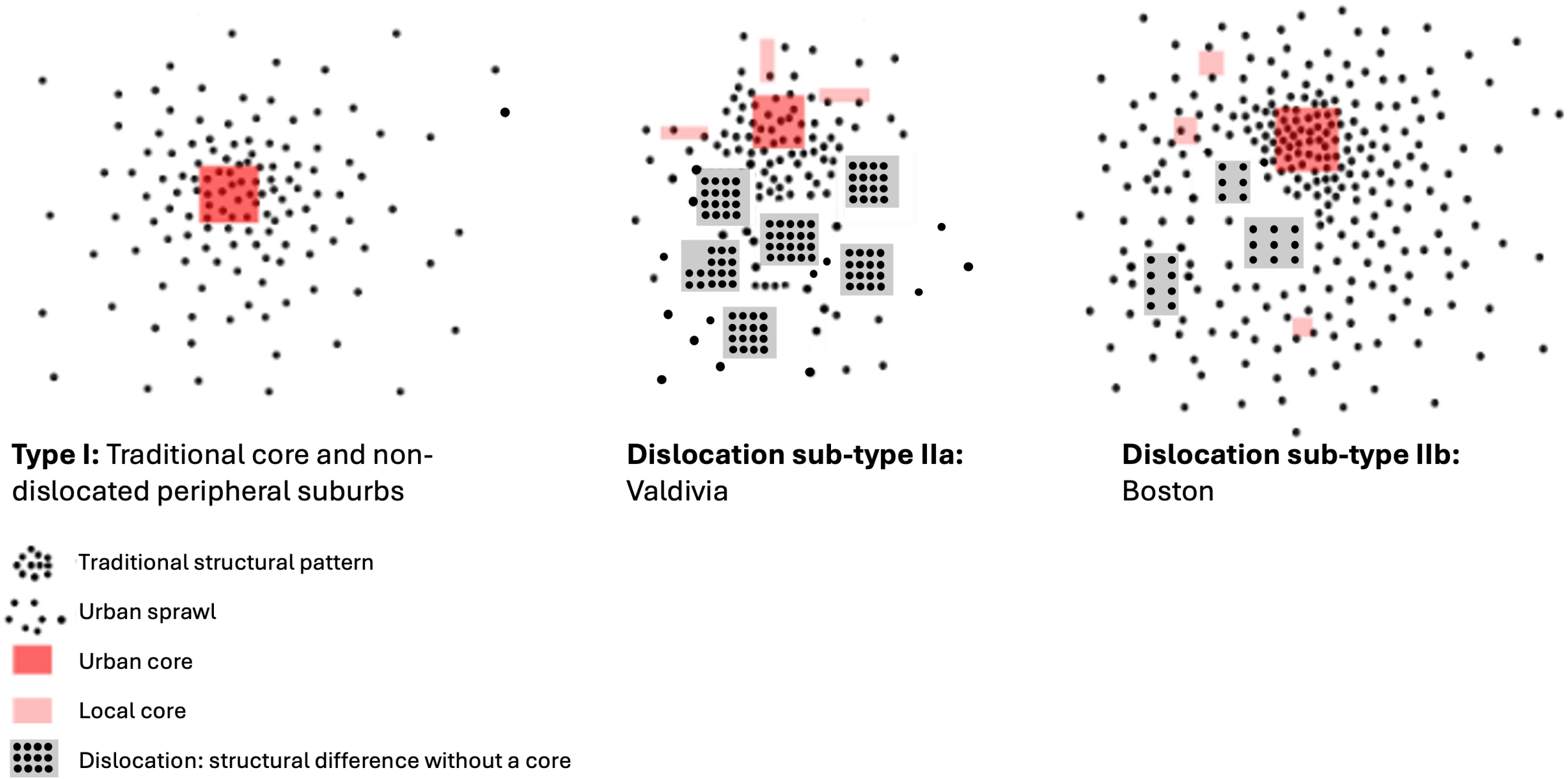

**Figure 5.** Typology of peripheral development. Type I: Traditional urban core surrounded by sprawled, non-dislocated development. Type II: Peri-urban dislocation, differentiating structurally from the core, with two subtypes: IIa, a peripheral reversal in which the peripheral areas cluster independently, at lower distances than the core and the rest of the system, and IIb, disconnected peri-urban voids, that remain segregated, and do not cluster with the rest of the system.

These typologies clarify why some peripheral growth integrates while other growth structurally misfits, helping distinguish dislocated from non-dislocated suburbs and from generic sprawl.

### 5.5. Limitations and future research

Our approach privileges the street network as the primary observation element, reflecting the method's foundation in physical patterns. While this reveals structural integration at the morphological level, it does not directly capture commuting patterns, economic flows, or social interactions. Additionally, our analysis treats the street network as a static structure at a single point in time, whereas dislocation patterns may evolve as infrastructure develops or areas consolidate. This constitutes a limitation that should be addressed in future developments.

Additionally, our empirical validation is limited to two contrasting cases: a mid-sized Chilean city and a large North American metropolis. While these cases demonstrate distinct manifestations of peri-urban dislocation, broader validation is needed across different geographies, city sizes, planning regimes, and development contexts.

Finally, while the method is adaptable, identifying dislocation requires some contextual understanding of the urban structure to visualise the problem. The exploration of the method in other cities would need to be compared to local morphological visualisations and historical development patterns.

## 6. Conclusion

We introduce peri-urban dislocation as a structural condition that captures hierarchical misalignments between inner-city and peripheral areas. Using percolation analysis on street networks, we revealed a structural reversal in Valdivia, where peripheral clusters dominate and overpower the traditional core, and peri-urban voids in Boston, where isolated clusters persist despite metropolitan consolidation. These patterns reflect fundamental disruptions in hierarchical coherence that are not reducible to density, land use, or peripherality alone.

Methodologically, we extend percolation analysis to diagnose core–periphery structural breaks, an application not previously explored. By integrating clustering maps, dendrograms, and LCC charts, the approach provides reproducible diagnostics that conventional sprawl metrics overlook, bridging urban morphology and complexity science.

Conceptually, we demonstrate that structural dislocation is a measurable condition. The proposed typology distinguishes dislocated from non-dislocated peripheral development, identifying which peripheral growth integrates hierarchically and which structurally misfits**.** This matters for planning: misaligned areas may require interventions prioritising structural connectivity and the creation of local centres over density increases alone.

By revealing hierarchical misalignments as quantifiable phenomena, this work opens directions for comparative urban analysis. The method is transferable across scales, inviting application to test whether the typology requires refinement and to explore how dislocation patterns emerge, persist, or resolve over time. Recognising peri-urban dislocation as a structural dimension of urban growth enriches both scholarly understanding and planning practice, offering diagnostic tools to address fundamental integration failures in metropolitan systems.

Future work could integrate land-use diversity, centrality measures, density, pedestrian/transport flows, and temporal evolution, testing if dislocation co-occurs with, follows, or substitutes classic sprawl, and whether targeted connectivity and centre-creation interventions could measurably reduce dislocation signatures over time. Longitudinal analysis tracking how dislocation patterns emerge, persist, or resolve as cities grow would be particularly valuable.

**Acknowledgements**

M.B.T. was supported by the National Agency for Research and Development of the Chilean Government (ANID) / Becas Chile Doctoral Scholarship, Grant No. 72190455.